\documentclass[12pt, preprint]{aastex}

\slugcomment{Accepted by {\it The Astrophysical Journal}}

\shorttitle{Spectral Breaks in Fermi Blazars}
\shortauthors{M. B\"ottcher}

\begin{document}

\title{VHE Gamma-Ray Induced Pair Cascades in Blazars and Radio Galaxies: 
Application to NGC 1275}

\author{P. Roustazadeh and M. B\"ottcher\altaffilmark{1}}

\altaffiltext{1}{Astrophysical Institute, Department of Physics and Astronomy, \\
Ohio University, Athens, OH 45701, USA}

\begin{abstract}
Recent blazar detections by HESS, MAGIC, and VERITAS suggest that
very-high-energy (VHE, $E > 100$~GeV) $\gamma$-rays may be produced 
in most, if not all, types of blazars, including those that possess 
intense circumnuclear radiation fields. In this paper, we investigate 
the interaction of nuclear VHE $\gamma$-rays with the circumnuclear
radiation fields through $\gamma\gamma$ absorption and pair production,
and the subsequent Compton-supported pair cascades. We have developed
a Monte-Carlo code to follow the spatial development of the cascade
in full 3-dimensional geometry, and calculate the radiative output
due to the cascade as a function of viewing angle with respect to 
the primary VHE $\gamma$-ray beam (presumably the jet axis of the 
blazar). We show that even for relatively weak magnetic fields, the
cascades can be efficiently isotropized, leading to substantial off-axis
cascade emission peaking in the {\it Fermi} energy range at detectable
levels for nearby radio galaxies. We demonstrate that this scenario
can explain the {\it Fermi} flux and spectrum of the radio galaxy
NGC~1275.
\end{abstract}

\section{Introduction}

Blazars, a class of active galactic nuclei (AGNs) comprised of
Flat-Spectrum Radio Quasars (FSRQs) and BL~Lac objects, exhibit 
some of the most violent high-energy phenomena observed in AGNs 
to date. Their spectral energy distributions (SEDs) are characterized 
by non-thermal continuum spectra with a broad low-frequency component 
in the radio -- UV or X-ray frequency range and a high-frequency 
component from X-rays to $\gamma$-rays. They show rapid variability
across the electromagnetic spectrum. In extreme cases, the very-high-energy
(VHE) $\gamma$-ray emission of blazars has been observed to vary on
time scales of just a few minutes \citep{albert07a,aharonian07}.

Leptonic and hadronic models \citep[for a recent review see, e.g.][]{boettcher07}
can generally successfully account for the overall SEDs observed from 
the known VHE $\gamma$-ray blazars, which are almost all BL Lac objects. 
BL Lac objects span a wide range of synchrotron peak frequencies, 
from IR to X-rays. According to the location of the synchrotron
peak, they are classified as Low frequency peaked BL Lacs (LBLs:
synchrotron peak in the IR), intermediate BL Lac objects (IBLs: 
synchrotron peak in the optical/UV) or High frequency peaked BL Lacs
(HBLs: synchrotron peak in the X-rays).
Until very recently, all known VHE $\gamma$-ray blazars were HBLs. 
However, the recent detections of the IBLs W~Comae, 3C~66A, and 
PKS~1424+240 by VERITAS \citep{acciari08,acciari09,acciari10}, the 
LBLs BL~Lacertae and S5~0716+714 by MAGIC \citep{albert07b,anderhub09}, 
and even the Flat-Spectrum Radio Quasars (FSRQs) 3C~279 by MAGIC 
\citep{albert08} and PKS~1510-089 by HESS \citep{wagner10} suggest
that most blazars are intrinsically emitters of VHE $\gamma$-rays.

The fact that most LBLs and FSRQs have not been detected as VHE $\gamma$-ray
sources might be primarily a result of the absorption of VHE $\gamma$-rays
by lower-frequency (IR -- optical -- UV) radiation. As those objects tend
to be located at greater cosmological distances ($z \gtrsim 0.2$), $\gamma\gamma$ 
absorption on the Extragalactic Background Light (EBL) becomes substantial,
suppressing any intrinsically produced $> 100$~GeV emission 
\citep[e.g.][]{dk04,stecker08,franceschini08,finke09,finke10}. 
However, possibly even more importantly, multi-GeV $\gamma$-rays produced
in the high-radiation-density environment within the broad line region
(BLR) and the dust torus of a quasar, is expected to be strongly attenuated 
by $\gamma\gamma$ pair production \citep[e.g.][]{pb97,donea03,reimer07,liu08,sb08}.
The studies cited above show that for typical parameters expected in 3C279, 
photons above $\sim 100$~GeV are expected to be strongly attenuated unless 
the $\gamma$-ray emission region is located beyond the broad line region. 
If VHE $\gamma$-rays are produced very close to the central engine,
also $\gamma\gamma$ absorption by the direct accretion disk emission may
become substantial \citep{sb10}.

Given the now substantial number of non-HBL VHE $\gamma$-ray blazars, it seems 
plausible
that most LBLs and FSRQs and their misaligned parent population, radio 
galaxies, are producing VHE $\gamma$-ray emission within their blazar zone.
This view is also supported by the detection of VHE $\gamma$-ray 
emission from two non-blazar AGNs, namely the radio galaxies M87
\citep{Aharonian04} and Cen~A \citep{Aharonian09}. 
The non-detection of VHE $\gamma$-rays from most LBLs and FSRQs
might then be due to the combined effects of local and intergalactic 
$\gamma\gamma$
absorption.
As a consequence of $\gamma\gamma$ absorption in the local
radiation field of the AGN, GeV -- TeV electron-positron pairs are injected 
into the AGN environment, which, in the dense radiation field within the 
BLR and the dust torus of quasars, may initiate inverse-Compton supported 
pair cascades. 

The development of pair cascades induced by VHE $\gamma$-ray emission
from blazars has so far concentrated on the development of Mpc-scale
pair halos resulting from the interaction of ($\gtrsim 100$~TeV) 
$\gamma$-rays with the Cosmic Microwave background 
\citep[e.g.][]{aharonian94}, or of VHE ($\gtrsim 100$~GeV) $\gamma$-rays
by the EBL \citep{venters10}. Due to the long Compton cooling timescale 
of pairs on the CMB or the EBL, those authors could reasonably 
consider the produced cascades isotropic,
leading to an extended pair 
halo around the AGN. 
Small-angle deflection of secondaries in weak intergalactic magnetic
fields has been included in those considerations by \cite{plaga95} and
\cite{elyiv09}. The development of pair cascades within the AGN has been
discussed by \cite{bk95} and \cite{sb10}. 

Depending on the magnetic
field strength and orientation in the extended 
nuclear region, cascades developing within the high-radiation-energy-density 
environment within the BLR of a
quasar may be efficiently isotropized in the 
immediate vicinity of the AGN. This will lead to distinct spectral features, 
which we will consider in this paper. In Section \ref{estimates} we present 
some general considerations and analytic estimates of the expected results, 
including an estimate of the required magnetic field strengths within $\sim$~a 
few pc from the central engine of a blazar (or radio galaxy), for which 
secondary electrons and positrons in pair cascades initiated 
by primary VHE $\gamma$-rays may be efficiently isotropized
within the central region. The consequence would be quasi-isotropic
radiation signatures from Compton-supported pair cascades, potentially
observable in the {\it Fermi} energy range. In Section \ref{setup} we will
describe a numerical code that treats the full three-dimensional development
of these cascades, together with the general model setup and simplifying
assumptions. Numerical results for generic parameters will be presented 
in Section \ref{parameterstudy}. In Section \ref{NGC}, we will
demonstrate that the recent {\it Fermi} detection of the radio galaxy
NGC~1275 \citep{abdo09a} can be plausibly explained by the cascade emission
from a misaligned VHE $\gamma$-ray emitting blazar. We summarize and present
an outlook towards future work in Section \ref{summary}.

\section{\label{estimates}General Considerations and Estimates}

In the limit of a VHE $\gamma$-ray photon with energy $E_{\gamma}$ 
interacting with an IR/optical/UV photon from the dust torus or the
BLR with energy $E_s \ll E_{\gamma}$, an electron-positron pair with
particle energy $E_e = \gamma \, m_e c^2 \approx E_{\gamma}/2 \equiv 
1 \, E_{\rm TeV}$~TeV will be produced, moving along the direction of the 
primary $\gamma$-ray photon to within an accuracy of $\Delta\theta \sim 
1/\gamma$. Depending on the strength of the magnetic field near the point of 
pair production, $B \equiv 1 \, B_{-6} \, \mu$G, they will be deflected on a
length scale of the order of the Larmor radius,
\begin{equation}
r_g \sim 10^{-3} \, E_{\rm TeV} \, B_{-6}^{-1} \; {\rm pc}.
\label{rg}
\end{equation}
To investigate whether particles are efficiently isotropized before
producing secondary synchrotron and/or inverse-Compton (IC) emission, 
the
isotropization length has to be compared with the radiative cooling
length, $\lambda_{\rm sy}$ and $\lambda_{\rm IC}$, respectively. 
The synchrotron cooling length can be estimated to
\begin{equation}
\lambda_{\rm sy} \sim 3.8 \times 10^6 \, E_{\rm TeV}^{-1} \, B_{-6}^{-2} \;
{\rm pc}.
\label{lambdasy}
\end{equation}
To estimate the IC cooling length on an external radiation 
field from the BLR we calculate its energy density as $u_{\rm BLR}
\sim L_D \tau_{\rm BLR} / (4 \pi R_{\rm BLR}^2 \, c)$, which we parameterize
through the accretion disk luminosity $L_D \equiv 10^{46} \, L_{46}$
erg~s$^{-1}$ and a BLR with a reprocessing optical depth of $\tau_{\rm
BLR} \equiv 0.1 \, \tau_{-1}$ at an average distance of $R_{\rm BLR}
\equiv 0.1 \, R_{-1}$~pc from the central engine. Hence, the total
luminosity of the BLR will be $L_{\rm BLR} = \tau_{\rm BLR} \, L_D$. 
If Compton scattering occurs in the Thomson regime, we find
\begin{equation}
\lambda_{\rm IC} \sim 5 \times 10^{-6} \, E_{\rm TeV}^{-1} \, L_{46}^{-1}
\tau_{-1}^{-1} R_{-1}^2 \; {\rm pc}.
\label{lambdaic}
\end{equation}
This illustrates that one may expect particles at energies substantially 
below 1~TeV to be fully isotropized, while higher-energy
particles might lose a substantial fraction of their energy while still
traveling along the primary VHE $\gamma$-ray beam. Comparison of the
synchrotron and IC cooling lengths suggests that the radiative output
from the secondaries will be strongly dominated by IC emission, initiating
an IC-supported cascade \citep[e.g.][]{protheroe86,zdziarski88}. At low
frequencies far below the pair production threshold, produced by secondaries
which are fully isotropized (i.e., $\lambda_{\rm IC} \gg r_g$), the cascade
spectrum will obtain a $\nu F_{\nu} \propto \nu^{1/2}$ shape. This is the
consequence of the secondaries being injected at high energies and then 
being subject to Compton cooling in the Thomson regime, resulting in a
$N(\gamma) \propto \gamma^{-2}$ pair spectrum. 
This low-energy spectral shape (though not its total flux) will be 
independent of the primary gamma-ray spectrum. 
Two effects will produce 
a turnover towards higher frequencies. First, for any given viewing angle 
$\theta$ with respect to the direction of propagation of the primary 
$\gamma$-ray (being absorbed in the $\gamma\gamma$ pair production process), 
we can find a critical electron energy for which the deflection angle over
a Compton length equals the observing angle, i.e., $\theta \sim \lambda_{\rm IC} 
/ r_g$. Higher-energy particles will radiate preferentially at smaller viewing
angles, while lower-energy particles can efficiently contribute to the emission
at the given angle. This yields the characteristic electron energy
$E_{\rm e, br}$ corresponding to a given observing angle $\theta$:

\begin{equation}
E_{\rm e, br} = m_e c^2 \, \sqrt{ 3 \, e \, B \over 4 \, \sigma_T \, u_{\rm BLR}
\, \theta} \; \sim \; 70 \, B_{-6}^{1/2} \, R_{-1} \, L_{46}^{-1/2}
\, \tau_{-1}^{-1/2} \, \theta^{-1/2} \; {\rm GeV}.
\label{Ebreak}
\end{equation}

If these electrons can scatter the soft photon field in the Thomson regime,
a turnover should occur at photon energies of

\begin{equation}
E_{\rm IC, br} = {3 \, e \, B \over 4 \, \sigma_T \, u_{\rm BLR} \, \theta} 
\, E_s \; \sim \; 18 \, B_{-6} \, R_{-1}^2 \, L_{46}^{-1} \, \tau_{-1}^{-1} \, 
\theta^{-1} \, \left( {E_s \over {\rm eV}} \right) \; {\rm GeV}.
\label{ICbreak}
\end{equation}

We point out that, in this estimate, we assumed that the Compton cooling length
is a realistic measure of the distance travelled by the electron/positron since
its production from annihilation of a primary $\gamma$-ray. This approximation
will break down if the particles require multiple Compton scatterings to reach
the break energy corresponding to Eq. \ref{ICbreak}. 
This is due to both an increasing Compton cooling length and a decreasing
Larmor radius as the electron energy is reduced by previous scatterings.

Second, if Compton scattering to energies given by Eq. \ref{ICbreak} occurs in 
the Klein-Nishina regime (which is the case if $E_{\rm IC, br} \ge E_{\rm e, br}$)
a turnover is expected at the transition from Thomson to Klein-Nishina scattering
at

\begin{equation}
E_{\rm IC, KN} = 260 \left( {E_s \over {\rm eV}} \right)^{-1} \; {\rm GeV}.
\label{KNbreak}
\end{equation}

\begin{figure}[ht]
\begin{center}
\includegraphics[width=8cm]{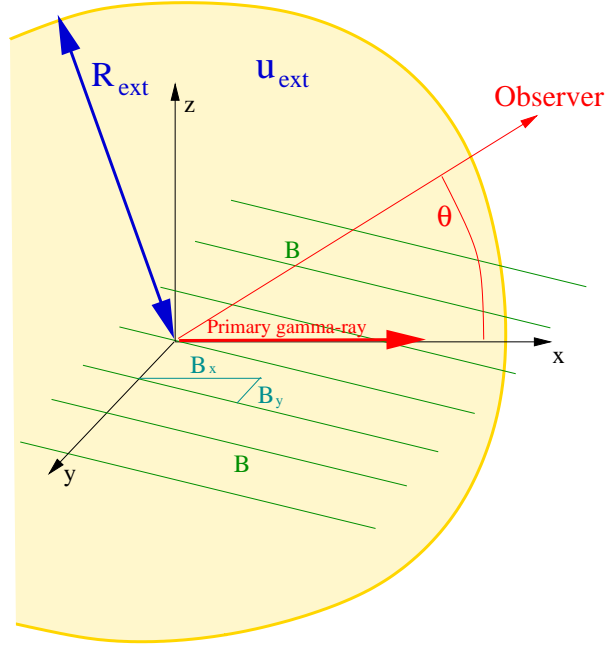}
\caption{\label{geometry}Geometry of the model setup.}
\end{center}
\end{figure}

\section{\label{setup}Model Setup and Code Description}

The geometrical setup of our model system is illustrated in Figure \ref{geometry}.
The primary VHE $\gamma$-ray emission from the blazar zone is represented as a
mono-directional beam of $\gamma$-rays propagating along the X axis. The incident 
VHE $\gamma$-ray spectrum is represented by a straight power-law with photon index 
$\alpha$. Those $\gamma$-rays may interact via $\gamma\gamma$ absorption and pair 
production with a radiation field. For a first general investigation and feasibility 
study presented here, we approximate the radiation field as monoenergetic and 
isotropic within a fixed boundary, given by a radius $R_{\rm ext}$, i.e.,

\begin{equation}
u_{\rm ext} (\epsilon, r, \Omega) = u_0 \, \delta(\epsilon - \epsilon_{\rm ext})
\, H(R_{\rm ext} - r)
\label{uapprox}
\end{equation}
where $H$ is the Heaviside function, $H(x) = 1$ if $x > 0$ and $H(x) = 0$ otherwise.
A magnetic field of order $\sim \mu$G is present. Without loss of generality, we
choose the $y$ and $z$ axes of our coordinate system such that the magnetic field
lies in the (x,y) plane. 

We have developed a Monte-Carlo code which treats the processes of $\gamma\gamma$
absorption and pair production, $\gamma$-ray and electron/positron propagation,
and Compton scattering. The code generates a single $\gamma$-ray photon at 
a time, at the origin of our coordinate system, propagating in the x 
direction. In order to improve the statistics of the otherwise very 
few highest-energy photons, we introduce a statistical weight inversely 
proportional to the photon energy. The code calculates the absorption 
opacity $\kappa_{\gamma\gamma}$, using the full pair production cross
section. Based on the corresponding absorption length, a location for the occurrence
of the next $\gamma\gamma$ absorption / pair production process is drawn. If the
pair production site is outside the radius $R_{\rm ext}$, the photon escapes; otherwise,
the photon is absorbed, and an electron-positron pair is created. The code uses the 
analytic result for the $\gamma\gamma$ pair production spectrum of \cite{bs97} to 
draw the energies of the electron and the positron. As motivated above, we assume
that both particles initially propagate in the direction of the absorbed $\gamma$-ray.
For both particles, the Compton scattering length $\lambda_{\rm IC}$ is calculated
using the full Compton cross section. As we expect the magnetic-field energy
density to be much smaller than the radiation energy density, we neglect synchrotron
losses to the electrons/positrons. Based on the value of $\lambda_{\rm IC}$, the
code draws a length that the electron/positron travels before the next Compton 
scattering event occurs. It then propagates the electron/positron, using the 
full 3-D geometry, through its gyrational motion in the magnetic field, to 
calculate the (x,y,z) coordinates and direction of motion of the 
electron/positron at the point of scattering. If this point of 
scattering is outside the radius $R_{\rm ext}$, the electron/positron
escapes; otherwise, Compton scattering occurs. The energy of the produced 
photon is drawn using a $\delta(\Omega_{\rm sc} - \Omega_e)$ approximation 
for the Compton cross section \citep[e.g.,][]{db06}, i.e., the scattered photon
is traveling in the same direction as the electron/positron before scattering. 
The produced ($\gamma$-ray) photon is then tracked through the same photon
tracking routine as the primary VHE $\gamma$-ray photons (properly accounting
for the correct location and direction of propagation). The energy of the
electron/positron is reduced by the energy of the scattered photon, and the
particle is returned to the pair tracking routine. If the energy of the
electron/positron is below a set threshold (determined by the condition that
they will no longer produce Compton emission in the energy range of interest),
the code will move on to the next particle. 

The energies, statistical weights, and directions of propagation of photons 
escaping from the region of high external radiation field (bounded by
$R_{\rm ext}$), are written into a photon event file. In a post-processing 
routine, this event file will be read to produce photon spectra with arbitrary 
energy and angular binning.

\begin{figure}[ht]
\begin{center}
\includegraphics[width=12cm]{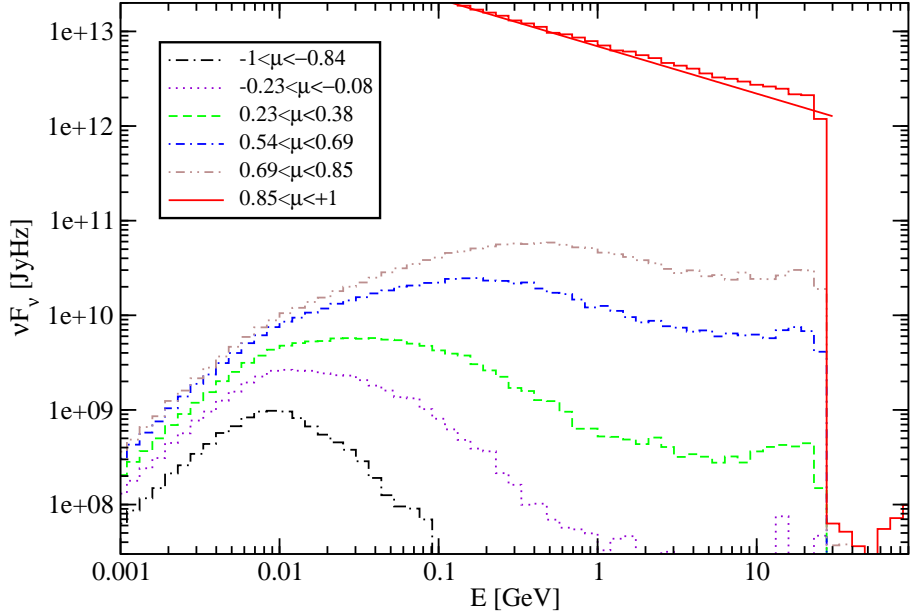}
\caption{\label{standardfig}
Cascade emission at different viewing angles 
($\mu = \cos\theta_{\rm obs}$). Parameters: $B = 1 \, \mu$G, $\theta_B = 5.7^o$; 
$u_0 = 10^{-2}$~erg~cm$^{-3}$, $R_{\rm ext} = 10^{19}$~cm, $E_s = 
E_{{\rm Ly}\alpha}$, $\alpha = 2.5$, {$E_{\gamma, {\rm max}} = 5$~TeV}.
The solid (red) straight line indicates the incident primary $\gamma$-ray 
spectrum.
}
\end{center}
\end{figure}

\section{\label{parameterstudy}Numerical Results}

We have used the cascade Monte-Carlo code described in the previous section
to evaluate the angle-dependent cascade spectra for a variety of generic
parameter choices. Figure \ref{standardfig} illustrates the viewing angle 
dependence of the cascade emission. For this simulation, we assumed a 
magnetic field of $B = 1 \, \mu$G, oriented at an angle $\theta_B = 5.7^o$ 
with respect to the X axis ($B_x = 1 \, \mu$G, $B_y = 0.1 \, \mu$G). The 
external radiation energy density is $u_0 = 10^{-2}$~erg~cm$^{-3}$, extended 
over a region of radius $R_{\rm ext} = 10^{19}$~cm with photon energy $E_s = 
E_{{\rm Ly}\alpha}$. The incident $\gamma$-ray spectrum has a photon index of
$\alpha = 2.5$ 
and extends out to $E_{\gamma, {\rm max}} = 5$~TeV. 
The results have been normalized to a flux level in the 
forward direction corresponding to a $\gamma$-ray bright blazar. The spectra
for all other directions have been normalized with the same normalization
factor. The curves are labeled by the cosine of the observing angle,
$\mu = \cos\theta_{\rm obs}$. In the forward (blazar) direction, one clearly
sees the $\gamma\gamma$ absorption cut-off at an energy $E_c = (m_e c^2)^2
/ E_s \sim 25$~GeV. 
The cutoff is very sharp in this simulation because of our $\delta$ 
approximation of the external radiation field, combined with a high $\gamma\gamma$
absorption depth near threshold for the parameters chosen here.
Below this cutoff, the forward-component of the cascade
emission leads to a slight bump beyond the primary $\gamma$-ray power-law
spectrum. The cascade emission at larger viewing angles $\mu < 0.85$ has 
a low-frequency shape close to the expected $\nu F_{\nu} \propto \nu^{1/2}$ 
behaviour, and exhibits
the progressive suppression of the cascade emission at high energies with 
increasing viewing angle due to incomplete isotropization of the secondary 
particles at high energies, as expected from Eq. \ref{ICbreak}.

\begin{figure}[ht]
\begin{center}
\includegraphics[width=12cm]{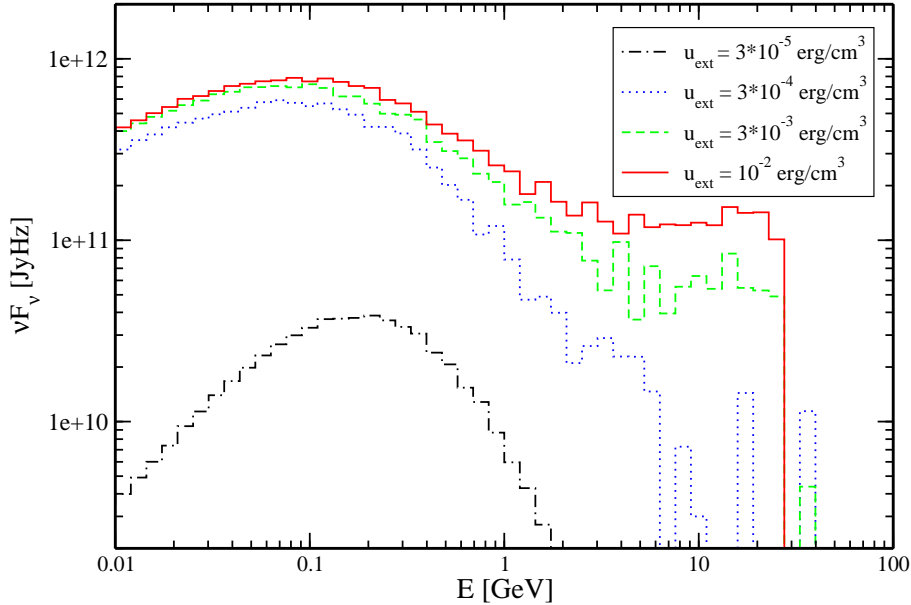}
\caption{\label{ufig}The effect of a varying external radiation energy density. 
Parameters are the same as for Figure \ref{standardfig} in the angular bin 
$0.38 \le \mu \le 0.54$. }
\end{center}
\end{figure}

Figure \ref{ufig} illustrates the effect of a varying external radiation field
energy density $u_0$. For energy densities $u_0 \gtrsim 10^{-3}$~erg~cm$^{-3}$, 
$\gamma\gamma$ absorption is essentially saturated, i.e., all VHE photons above 
the pair production threshold will be absorbed. Hence, the magnitude of the cascade 
becomes almost independent of $u_0$. For smaller $u_0$, the decreasing flux in the
cascade emission reflects the decreasing fraction of VHE $\gamma$-ray photons
absorbed. This latter is the regime in which our estimate of the turnover
frequency (Eq. \ref{ICbreak}) is applicable. In the saturated regime, the
electrons/positrons have to undergo many scatterings before reaching
the isotropization energy so that the Compton scattering length is no
longer an appropriate measure of the distance traveled, as assumed in 
the derivation of Eq. \ref{ICbreak}. As expected, the low $u_0$ case
results in a larger turnover energy than the high $u_0$ cases.

\begin{figure}[ht]
\begin{center}
\includegraphics[width=12cm]{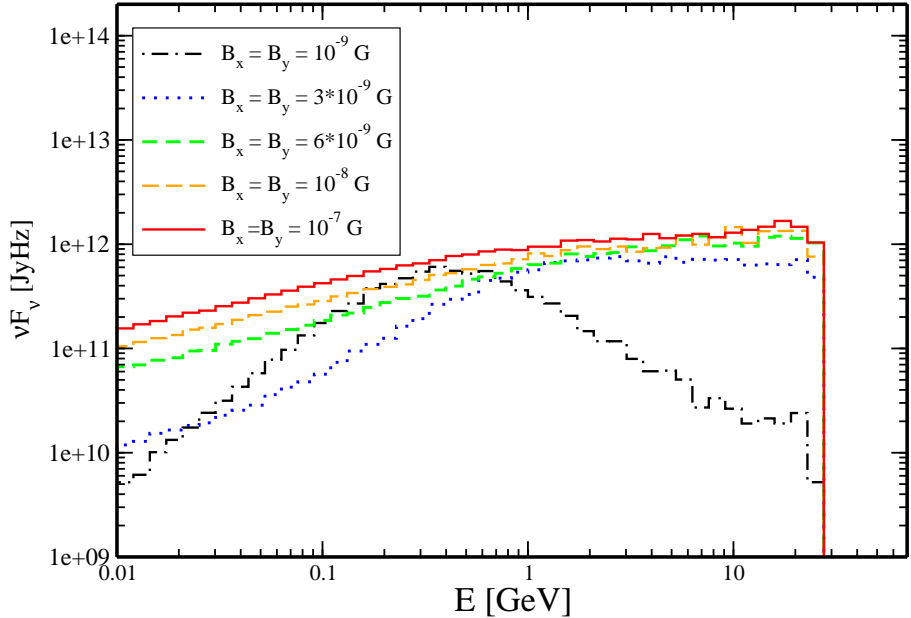}
\caption{\label{Bfig}The effect of a varying magnetic field strength, for a 
fixed angle of $\theta_B = 45^o$ between jet axis and magnetic field. All 
other parameters are the same as for Figure \ref{standardfig} in the angular 
bin $0.38 \le \mu \le 0.54$. }
\end{center}
\end{figure}

The effect of a varying magnetic field strengh for fixed magnetic field 
orientation ($\theta_B = 45^o$) is illustrated in Figure \ref{Bfig}.
We see that the cascade development is extremely sensitive to the 
transverse magnetic field $B_y$ for weak magnetic fields. The cascades
exhibit a rapid transition to the limit in which even the highest-energy 
secondary particles are effectively isotropized before undergoing the 
first Compton scattering interaction. Hence, for magnetic field values
expected on sub-pc or pc scales around an AGN ($B \gg 1$~nG) and large 
angles $\theta_B$, there is no pronounced break in the cascade spectrum 
out to large energies near the $\gamma\gamma$ absorption trough at $E_c$.

\begin{figure}[ht]
\begin{center}
\includegraphics[width=12cm]{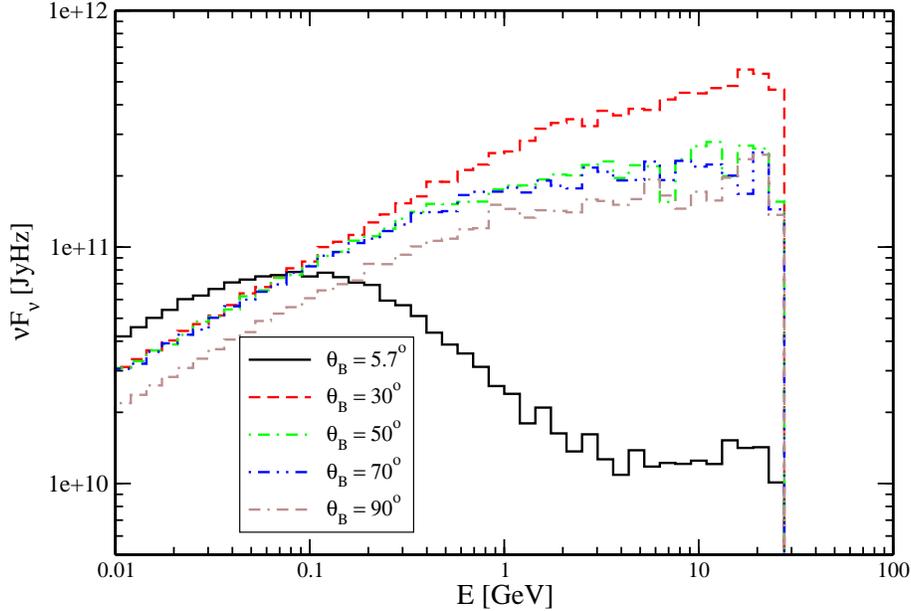}
\caption{\label{Bthetafig}The effect of a varying magnetic field orientation, 
for a fixed magnetic field strength of $B = 1 \, \mu$G. All other parameters 
are the same as for Figure \ref{standardfig} in the angular bin $0.38 \le \mu 
\le 0.54$.}
\end{center}
\end{figure}

Figure \ref{Bthetafig} shows the effect of a varying magnetic-field orientation 
with respect to the jet axis, for fixed magnetic-field strength $B = 1 \, \mu$G. 
The figure illustrates that it is primarily the perpendicular ($B_y$) component 
of the magnetic field which is responsible for the isotropization of secondaries 
in the cascade. Obviously, for a perfectly aligned magnetic field ($\theta_B = 0^o$),
our code does not predict any cascade emission in any off-axis direction, since we
neglect the spreading of the cascade due to the kinematics of the pair-production
process or recoil from Compton scattering. For a small inclination angle, the
small perpendicular magnetic-field component leads to inefficient isotropization
and, hence, a break in the off-axis cascade spectrum at low energies. For large
inclination angles ($B_y \gtrsim B_x$), isotropization becomes very efficient 
out to energies close to the $\gamma\gamma$ absorption trough.

\begin{figure}[ht]
\begin{center}
\includegraphics[width=12cm]{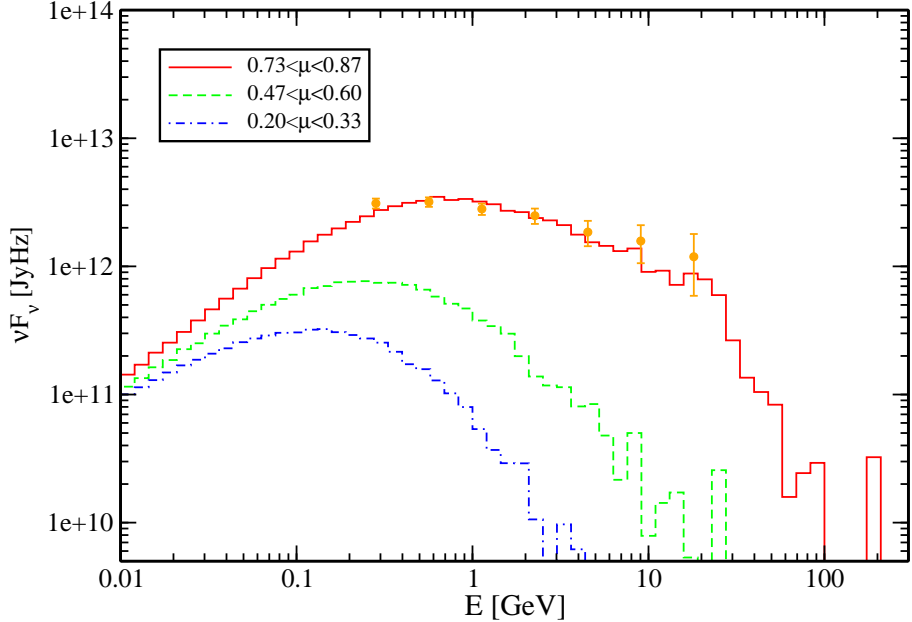}
\caption{\label{fit}Fit to the {\it Fermi} spectrum of NGC~1275 with a 
simulated cascade spectrum from a mis-aligned blazar, along with the 
cascade spectra at larger viewing angles. }
\end{center}
\end{figure}

\section{\label{NGC}Application to Radio galaxies: The case of NGC~1275}

From the numerical results shown above it has become obvious that VHE $\gamma$-ray
induced cascades even in the presence of rather weak ($\sim \mu$G) magnetic fields
can be efficiently isotropized and produce a substantial MeV -- GeV $\gamma$-ray
flux in directions misaligned with respect to the jet axis (the forward or ``blazar''
direction). The standard AGN unification scheme \citep{up95} proposes that blazars
and radio galaxies are intrinsically identical objects viewed at different angles
with respect to the jet axis. According to this scheme, FR~I and FR~II radio galaxies 
are believed to be the parent population of BL~Lac objects and FSRQs, respectively.
Hence, if most blazars, including LBLs and FSRQs, are intrinsically VHE $\gamma$-ray
emitters potentially producing pair cascades in their immediate environments, the
radiative signatures of these cascades might be observable in many radio galaxies.
In fact, already three radio galaxies (NGC~1275: \cite{abdo09a}, M~87: \cite{abdo09b}, 
and Cen~A: \cite{abdo09c}) have been detected by Fermi, while {\it EGRET} provided
evidence for $> 100$~MeV $\gamma$-ray emission from two more radio galaxies, 
3C~111 \citep{nandi07} and NGC~6251 \citep{muk02}. In this paper, we focus on
the radio galaxy NGC~1275 \citep{abdo09a} and investigate whether the Fermi spectrum
and flux level are consistent with the interpretation of cascade emission from a 
misaligned, $\gamma\gamma$ absorbed blazar. 

NGC~1275 is located at a distance of $d = 74$~Mpc. It hosts the powerful FR~I radio 
galaxy Perseus A (= 3C~84). The radio jet is directed at an angle of $\theta \approx 
30^o$ -- $55^o$ with respect to the line of sight \citep{walker94}. 
NGC~1275 also hosts a Seyfert-like nucleus with a total line luminosity 
of $L_{\rm BLR} = 1.6 \times 10^{42}$~erg~s$^{-1}$ \citep{zb95}. Fermi 
observed a $\gamma$-ray flux of $F(> 100 \, {\rm MeV}) = (2.10 \pm 0.23) \times 
10^{-8}$~ph~cm$^{-2}$~s$^{-1}$ from NGC~1275, which could be fit by a power-law with a 
photon index of $\Gamma = 2.17 \pm 0.05$ \citep{abdo09a}. However, the Fermi
spectrum does seem to exhibit a marked steepening towards higher energies. 
A comparison with the EGRET upper limit  indicates substantial flux variability 
on time scales of years to decades, indicating that the $\gamma$-ray emission 
must be produced in a region of $R \lesssim$~a few pc. 

Our simulation results can be properly normalized to the expected flux for a
radio galaxy with the following considerations: The forward spectrum shown 
in Figure \ref{standardfig} corresponds to the escaping radiation
in a solid angle interval $\Delta\Omega = 2 \pi \, \Delta\mu$, where $\Delta\mu
= 0.15$ in the examples shown here. However, the cone of blazar emission might
be smaller, namely $\Delta\Omega_b \approx \pi \theta_b^2$, where $\theta_b$
is the opening angle of the blazar emission cone. Hence, a realistic flux
normalization for any off-axis direction requires a correction factor of
$f_{\rm ang} = \theta_b^2 / (2 \, \Delta\mu)$. 
The normalization will also depend on
the difference in distances, $f_{\rm dist} = (d_{\rm bl}
/ d_{\rm RG})^2$, where $d_{\rm bl}$ and $d_{\rm RG}$ are the luminosity 
distances of the blazar and the radio galaxy, respectively. 
Hence, we apply a total normalization factor $f = f_{\rm ang} \, 
f_{\rm dist}$.

Figure \ref{fit} illustrates that the Fermi spectrum of NGC~1275 can be well
matched with a VHE $\gamma$-ray induced cascade from a misaligned blazar. For
the normalization of the incident $\gamma$-ray beam, we used the Fermi spectrum
of the blazar 3C279 at $d_L = 2.2$~Gpc. The external radiation field is
parameterized through $u_{\rm ext} = 5 \times 10^{-2}$~erg~cm$^{-3}$ and 
$R_{\rm ext} = 10^{16}$~cm. This size scale is appropriate for low-luminosity 
AGN as observed in NGC~1275 \citep[e.g.][]{kaspi07}, and the parameters combine 
to a BLR luminosity of $L_{\rm BLR} = 4 \pi R_{\rm ext}^2 \, c \, u_{\rm ext} 
= 1.9 \times 10^{42}$~erg~s$^{-1}$, in agreement with the observed value for 
NGC~1275. The magnetic field is $B = 1$~mG, oriented at an angle of
$\theta_B = 8^o$. The cascade spectrum shown in Figure \ref{fit} pertains 
to the angular bin $0.73 < \mu < 0.87$ (corresponding to $30^o \lesssim 
\theta \lesssim 43^o$), appropriate for the known orientation of NGC~1275. 
The factor required to re-normalize between the simulated spectrum in the 
forward direction (at a flux level observed for 3C~279) and the sideways 
spectrum, fit to NGC~1275, was $f = f_{\rm ang} \, f_{\rm dist} = 146$. 
This corresponds to a blazar cone opening angle of $\theta_b = 12.7^o$ 
for 3C279. It should be pointed out that our normalization to flux levels of
3C279 are only meant to demonstrate that our parameter values correspond 
to conditions reasonably expected in misaligned blazars in general. The
numbers quoted above pertaining to 3C~279 should not be taken at face value
as a diagnostic for 3C279 itself.

\section{\label{summary}Summary and Outlook}

We have developed a Monte-Carlo code to simulate the full 3-D development of
VHE $\gamma$-ray induced pair cascades in the weakly magnetized environments 
of blazars and radio galaxies. These cascades develop as Compton-supported
pair cascades if there is a substantial radiation energy density, e.g., from 
the broad-line region or the dust torus around the central engine. We have
demonstrated the angle-dependence of these cascades as a function of the 
magnetic field and the radiation energy density in the AGN environment. In
the off-axis directions, these cascades exhibit a $\nu F_{\nu} \propto 
\nu^{1/2}$ spectrum at low energies and a turnover towards higher energies 
due to incomplete isotropization of high-energy secondares in the cascade,
depending on the strength and orientation of the magnetic field and the 
radiation energy density.

We have demonstrated that the off-axis cascade emission may well be detectable
from nearby radio galaxies, which, according to the blazar unification scheme,
are the mis-aligned parent population of blazars. We have presented a fit 
to the {\it Fermi} spectrum of the radio galaxy NGC~1275 with parameters
expected if this radio galaxy is the mis-aligned counterpart of a typical
{\it Fermi}-detected blazar. 

The present work is to be understood as a general proof of the potential importance
of VHE $\gamma$-ray induced pair cascades in blazar environments. For this purpose,
we have used a simplified representation of the radiation field as monoenergetic, 
homogeneous and isotropic out to a limiting radius. In future work, we will relax 
this assumption to allow for an arbitrary radiation spectrum and more realistic 
spatial and angular dependence. While we expect slight quantitative differences 
in the spectral shape of the emanating cascade emission due to these modifications, 
the general viability of the process as well as its dependence on magnetic field
properties and the external radiation field are expected to remain robust 
predictions of our work.

\acknowledgements{We thank Jun Kataoka for sending us the Fermi data points for
NGC~1275. This work was supported by NASA through Fermi Guest Investigator Grant 
NNX09AT81G.}

\end{document}